\documentstyle[aps,twocolumn,prl,epsf]{revtex}

\begin{document}
 
\draft

\title{Microextensive Chaos of a Spatially Extended System}

\author{
Shigeyuki Tajima\cite{CNCS-address}\cite{tajima-email} and
Henry~S. Greenside\cite{CNCS-address}
}

\address{
Department of Physics\\
Duke University, Durham, NC 27708-0305
}

\date{June 4, 2001}

\maketitle

\begin{abstract}
  By analyzing chaotic states of the one-dimensional
  Kuramoto-Sivashinsky equation for system sizes~$L$ in
  the range~$79 \le L \le 93$, we show that the
  Lyapunov fractal dimension~$D$ scales {\em
    microextensively}, increasing linearly with~$L$
  even for increments~$\Delta{L}$ that are small
  compared to the average cell size of~$9$ and to
  various correlation lengths. This suggests that a
  spatially homogeneous chaotic system does not have to
  increase its size by some characteristic amount to
  increase its dynamical complexity, nor is the
  increase in dimension related to the increase in the
  number of linearly unstable modes.
\end{abstract}

\vspace{.2in}

\pacs{
05.45.Jn,   
05.45.-a,   
47.27.Eq  
}

\narrowtext

An important phenomenon associated with sustained
nonequilibrium systems is spatiotemporal chaos, a
chaotic dynamical state that is spatially
disordered~\cite{CrossHohenberg94,Cross93}.  An open
question is how best to characterize spatiotemporal
chaos so that theory can be quantitatively compared
with experiment and experiment with simulation.
Presently, there is no fundamental theory of
nonequilibrium systems to indicate the appropriate
quantities to measure and so researchers have borrowed
ideas from condensed matter physics, fluid dynamics,
nonlinear dynamics, and statistics.  Commonly used ways
to characterize spatiotemporal chaos include critical
exponents~\cite{Marcq97}, the two-point correlation
time~$\tau_2$~\cite{Egolf94spatial}, the largest
Lyapunov exponent~$\lambda_1$~\cite{Cross93}, the
Lyapunov fractal dimension~$D$, the two-point
correlation length~$\xi_2$~\cite{Gluckman95}, the
dimension length~$\xi_\delta$
\cite{Cross93,Greenside99Montreal} and other
lengths~\cite{Cross93,OHern96,Zoldi98pre,Wittenberg99}.
However, calculations have shown that these quantities
do not always lead to the same conclusions, e.g., there
are systems for which the length~$\xi_2$ diverges while
the length~$\xi_\delta$ remains finite as some
parameter is varied~\cite{OHern96}. Further research is
therefore needed to understand the particular features
of spatiotemporal chaos that are measured by any one of
these quantities and how these quantities are related
to one another.

In the following, we report results that provide new
insights about how the dynamical complexity of a
nonequilibrium system depends on the volume of the
system, and about the interpretation of the dimension
length~$\xi_\delta$.  In 1982, Ruelle
conjectured~\cite{Ruelle82}, and numerical calculations
later
confirmed~\cite{Manneville85,OHern96,Strain98prl,Xi00pre,Egolf00Nature},
that the dimension~$D$ of a sufficiently large {\em
  spatially homogeneous} chaotic system should increase
extensively, i.e., linearly with its volume~$V$. Using
an argument similar to that used by Landau and Lifshitz
to explain the extensivity of additive quantities in
thermodynamics~\cite{Landau80}, this extensivity of~$D$
can be understood heuristically as a consequence of
spatiotemporal disorder. If two subsystems of a
spatiotemporal chaotic system are sufficiently far
apart, their coupling is weak because of the disorder
and so their dynamics contribute independently and
additively to the overall fractal dimension.

This picture of weakly interacting subsystems raises
the question of how precisely does the fractal
dimension~$D$ increase with increasing system volume in
the extensive regime. One possibility is that the
curve~$D(V)$ may be linear only on average and has a
staircase-like structure, with the steps corresponding
to new degrees of freedom that appear once the system
volume has increased sufficiently to include a new
subsystem.  The widths~$\Delta{V}$ of the steps would
then define a length scale~$(\Delta{V})^{1/d}$
(where~$d$ is the spatial dimensionality of the system)
that would be interesting to compare with the lengths
mentioned above ($\xi_2$, $\xi_\delta$, etc.).
Possible step-like features in the~$D(V)$ curve might
also be associated with the appearance of new linearly
unstable modes of the uniform state, since the number
of such modes typically increases linearly on average
with increasing volume~\cite{Pomeau84}.  Another
possibility is that the curve~$D(V)$ is extensive only
on average but its deviation from linearity is too
irregular to characterize by a single length scale.  A
fourth possibility is that there are no length scales
associated with how~$D$ increases with~$V$ and the
curve~$D(V)$ is exactly linear for arbitrarily small
increases in~$V$, a situation that one could call {\em
  microextensive chaos}. In this case, it would be
interesting to understand how the geometric structure
of the chaotic attractor in phase space changes
with~$V$ so as to produce such an exact linear
behavior.

In this paper, we numerically integrate the
one-dimensional Kuramoto-Sivashinsky (KS) equation---a
widely studied continuum model of spatiotemporal
chaos~\cite{Cross93}---to investigate how the Lyapunov
fractal dimension~$D(L)$ of a homogeneous chaotic
system varies with the system size~$L$ for
increments~$\Delta{L}$ that are small compared to the
lengths mentioned above ($\xi_2$, $\xi_\delta$, etc).
With one exception~\cite{Xi00pre}, all prior numerical
studies used increments~$\Delta{L}$ that were large
compared to these lengths and the detailed form
of~$D(L)$ was not determined. We show below that, in
fact, the Lyapunov fractal dimension~$D$ increases
linearly with~$L$ even for system
increments~$\Delta{L}$ that are tiny compared to the
average cell size and to various correlation lengths.
The spatiotemporal dynamics of the one-dimensional KS
equation therefore provides an example of
microextensive chaos. We conjecture that this will be a
general property of chaotic homogeneous nonequilibria
media.

Our calculations of~$D$ versus~$L$ yield an additional
insight, namely that the onset of extensivity in~$D$ is
not sharp but occurs only asymptotically with
increasing~$L$, after a sequence of alternating windows
of stationary, periodic, intermittent, and chaotic
dynamics. (Such alternating windows have been noted
before~\cite{Manneville88} but have not been studied
with such fine resolution in~$L$ as we do here.)  These
results suggest the possibility that windows of
non-chaotic behavior may persist to arbitrarily large
values of~$L$ but become too narrow to be detected. If
true, then the dimension~$D(L)$ may not be a continuous
curve and extensive behavior occurs only between the
narrow windows of non-chaotic dynamics.

\begin{figure}   
  \centerline{\epsfysize=3in \epsfbox{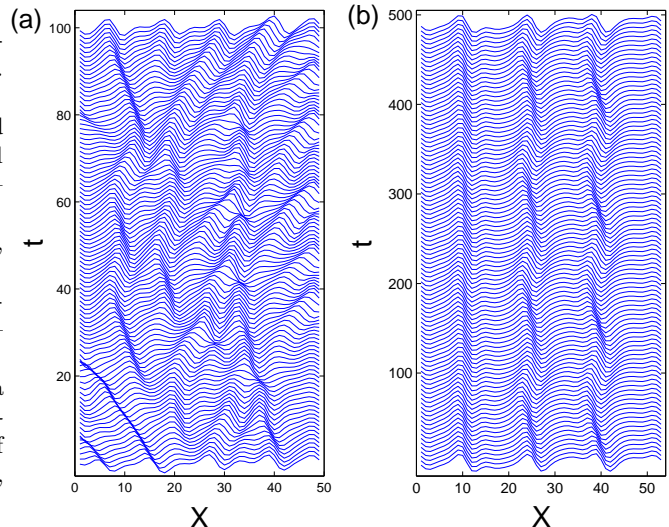}}
  \caption{ 
    Space-time evolution of the field~$u(t,x)$ for two
    states of the Kuramoto-Sivashinsky equation
    Eq.~(\ref{eq:ks-eq}) with rigid boundary
    conditions.  The space-time resolution
    was~$\Delta{t}=0.025$ and~$\Delta{x}=0.166$ and the
    peak-to-peak amplitude is about~4. {\bf (a):}
    Chaotic state for~$L=50$.  Spatial curves are
    plotted every~$\Delta{T}=1$ time units starting at
    time~$t=50,000$. {\bf (b)}: A periodic state
    for~$L=54$ with period~$\tau=127.6$.  Spatial
    curves are plotted every~$\Delta{T}=5$ time units
    starting at time~$t=80,000$.}
  \label{fig:ks-fields}
\end{figure}

Our results were obtained by numerical integrations of
the one-dimensional Kuramoto-Sivashinsky equation in
the form
\begin{equation}
  \partial_t u(t,x) = 
    - \partial_x^2 u
    - \partial_x^4 u
    - u \partial_x u
  , \qquad x \in [0,L]
  , \label{eq:ks-eq}
\end{equation}
on an interval of length~$L$, with rigid boundary
conditions $u = \partial_x u = 0$ at $x=0$ and
at~$x=L$.  (Fig.~\ref{fig:ks-fields} shows a chaotic
and periodic state for~$L=50$ and~$54$ respectively.)
The spatial derivatives were approximated by
second-order-accurate finite differences on a uniform
spatial mesh, and a standard operator-splitting method
was used for the time integration~\cite{Manneville90}.
For given initial conditions and interval length~$L$,
we used the Kaplan-Yorke formula~\cite{Parker89} to
calculate the Lyapunov fractal dimension~$D(L)$ in
terms of all of the positive and some of the negative
Lyapunov exponents~$\lambda_i$. These exponents were
obtained using a standard algorithm~\cite{Parker89} in
which many copies of the linearized KS~equation were
integrated, each with their own initial condition.

The demonstration of microextensive scaling by the
above numerical methods was delicate since the Lyapunov
exponents~$\lambda_i$ and so~$D$ converge noisily and
slowly~\cite{Goldhirsch87} toward their infinite time
limits.  As the increment~$\Delta{L}$ in system size
became smaller, the corresponding increment in
dimension~$\Delta{D}$ was more difficult to determine
since~$\Delta{D}$ became comparable to the fluctuations
in the dimension curve~$D(T)$ as a function of
integration time. The exponents~$\lambda_i$ and
dimension~$D(L)$ were also sensitive to the values of
the spatial resolution~$\Delta{x}$ and temporal
resolution~$\Delta{t}$, to the renormalization
time~$T_{\rm norm}$ for the Lyapunov vectors, and to
the total integration time. For nearly all runs
reported below, we used values of $\Delta{x}=0.167$,
$\Delta{t}=0.025$, and~$T_{\rm norm}=10$ and confirmed
the correctness of the corresponding results by
comparing the values with spatial and temporal
resolutions up to four times larger and for integration
times as long as~$10^6$~time units.

We now turn to our results. Our starting point was the
pioneering calculation of
Manneville~\cite{Manneville85}, who used numerical
integrations of Eq.~(\ref{eq:ks-eq}) with rigid
boundary conditions to demonstrate for the first time
that the fractal dimension~$D$ scaled extensively with
the system size~$L$. For~$L \ge 50$, he found that $D =
0.230L - 2.70$, which implies a dimension
length~\cite{Greenside99Montreal} of $\xi_\delta =
(dD/dL)^{-1} = 1/0.230 \approx
4.4$~\cite{other-growth-rates}.  This length is
%
%
somewhat smaller than the average cellular size
$\lambda = 2\pi/q_{\rm max} = 2\sqrt{2}\pi \approx 8.8$
corresponding to the fastest growing linear
mode~$q_{\rm max}=1/\sqrt{2}$. Based on these results,
we chose to calculate the fractal dimension~$D(L)$ over
the range~$50 < L < 100$ in constant
increments~$\Delta{L} = 0.5$ that were much smaller
than these lengths. In contrast, the smallest increment
used by Manneville was~$\Delta{L}=50$ for which the
fractal dimension changes by about~12.

\begin{figure}   
  \centerline{\epsfysize=3in \epsfbox{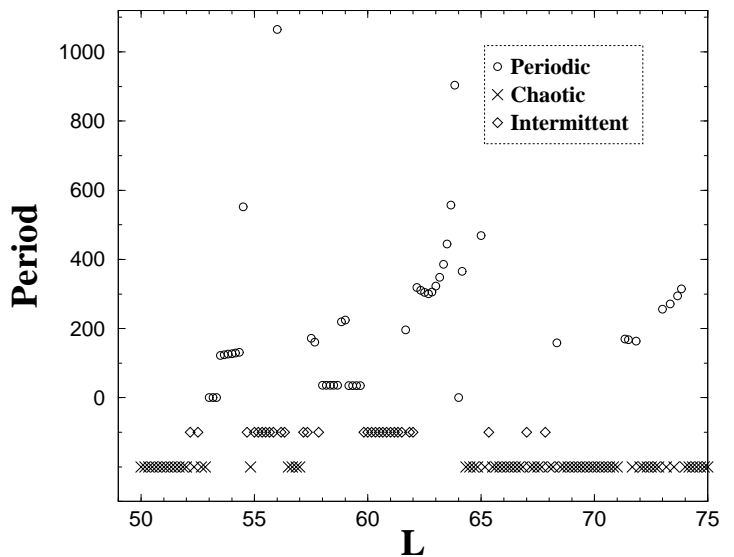}}
  \caption{
    Periods of numerical solutions to
    Eq.~(\ref{eq:ks-eq}) versus the system length~$L$.
    Each integration was started from small-amplitude
    random initial condition and then integrated
    500,000 time units. There is a complex sequence of
    windows corresponding to chaotic, constant,
    intermittent, and periodic dynamics.  Chaotic and
    intermittent solutions have been assigned an
    arbitrary period of -200 and -100 respectively so
    that all the data could be compared on one plot.  }
  \label{fig:ks_period}
\end{figure}

Manneville's linear dependence of~$D$ on~$L$ suggested
that for~$L \ge 50$, only spatiotemporal chaos exists.
In contrast, we find that there is a complicated
sequence of different dynamical states over the
range~$50 \le L \le 75$ and then only chaotic states
for~$75 < L < 93$~\cite{other-bifurcation-results}.
Fig.~\ref{fig:ks_period} summarizes our results for the
range~$50 \le L \le 75$ by plotting the period of each
state as a function of~$L$.  We observe four kinds of
states: fixed points, time-periodic states, chaos, and
intermittent states in which one kind of time
dependence alternates irregularly with a different kind
of time dependence. In most cases, these different
categories were easily identifiable to the eye by
looking at time series.  To combine all the results on
a single plot, chaotic states were arbitrarily assigned
a period of~-200, intermittent states a period of~-100,
fixed points a period of~0, and periodic states a
direct estimate of their period based on repeating
features of the time series.

There are two interesting features of the dynamical
states of Fig.~\ref{fig:ks_period} in addition to the
unexpected occurrence of many windows of alternating
dynamics for this range of~$L$.  First, we found that
for a given system size~$L$, numerical integrations
using up to seven different random initial conditions
(each consisting of uniformly distributed numbers in
the interval~$[-0.1,0.1]$) led to only one state. Thus
empirically there seems to be only one basin of
attraction for each system size and we do not expect
hysteresis in the range~$50 < L < 75$.  Second, we
found rather remarkably that the fractal dimension~$D$
of each chaotic state in Fig.~\ref{fig:ks_period} lay
on Manneville's extensive curve~$D(L)=0.230L-2.70$
with~$D \ge D(50) = 8.8$ (A least-squares fit of our
chaotic states gave the almost identical curve $0.227L
- 2.85$). Thus the states jump abruptly from
low-dimensional $D=1$ periodic states to
high-dimensional chaotic states that are scaling
extensively with the system size. We did not try to
characterize the intermittent states, e.g., by their
fractal dimension or by the scaling properties of the
fractional duration of a particular
phase~\cite{Berge84}.

\begin{figure}   
  \centerline{\epsfysize=3in \epsfbox{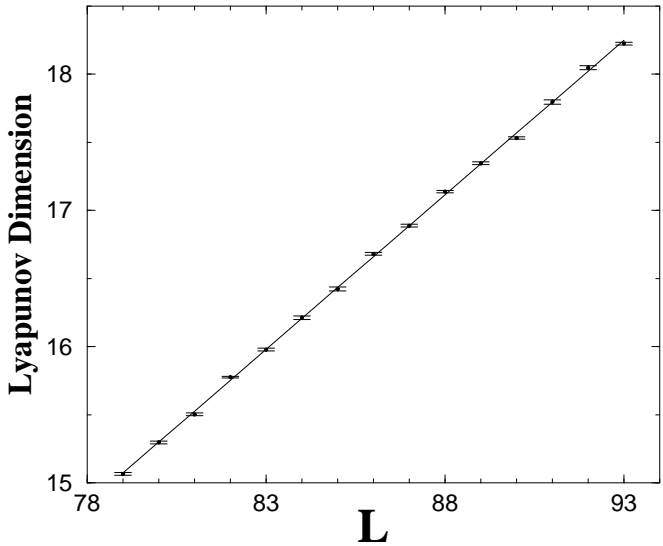}}
  \caption{
    The Lyapunov fractal dimension~$D$ of chaotic
    solutions to Eq.~(\ref{eq:ks-eq}) versus system
    size~$L$ for~$79 \le L \le 93$. The dimension
    values accurately fall on a straight line,
    demonstrating the occurence of microextensive
    scaling. The straight line $D = 0.227L -2.85$ was
    obtained by a least-squares fit to the points and
    agrees well with Manneville's
    result~\protect\cite{Manneville85} of~$D=0.230L -
    2.70$ over the much larger range $50 < L < 400$.
    The error bar for each point corresponds to a
    relative error of at most 0.05\% in~$D$.  The error
    bar was determined by the peak-to-peak fluctuations
    of~$D$ versus integration time~$T$.  }
  \label{fig:Lyapunov_dimension}
\end{figure}

Over the range $78 < L < 93$, only chaotic states were
observed. Fig.~\ref{fig:Lyapunov_dimension} shows that
the corresponding values of the Lyapunov fractal
dimension~$D$ lie on a straight line that has the same
slope (to two digits) and intercept as that found by
Manneville over the much larger range~$50 < L < 400$.
{\em Thus the fractal dimension shows microextensive
  scaling: a linear dependence on~$L$ for system
  increments~$\Delta{L}=0.5$ that are much smaller than
  any characteristic length scale such as the average
  cell size or various correlation lengths.} Given the
similar result obtained by Xi et al~\cite{Xi00pre} for
a different mathematical model of spatiotemporal chaos,
we conjecture that microextensive scaling will be a
general feature of spatiotemporal chaos in sufficiently
large, approximately homogeneous nonequilibrium
systems.

The linear behavior of
Fig.~\ref{fig:Lyapunov_dimension} rules out a simple
relation between the Lyapunov fractal dimension and the
number of linearly unstable modes. For both periodic
and rigid boundary conditions, the wave numbers are
quantized in units of~$2\pi/L$ and so the number of
unstable modes increases linearly as~$L/\pi \approx
0.32L$, but in discrete jumps when~$L$ changes by
about~$\pi$. Over the range $78 < L < 93$, we would
expect $(93-78)/\pi \approx 4$ new linearly unstable
modes to appear but there are not correspondingly four
step-like features in the figure.

\begin{figure}   
  \centerline{\epsfysize=3in \epsfbox{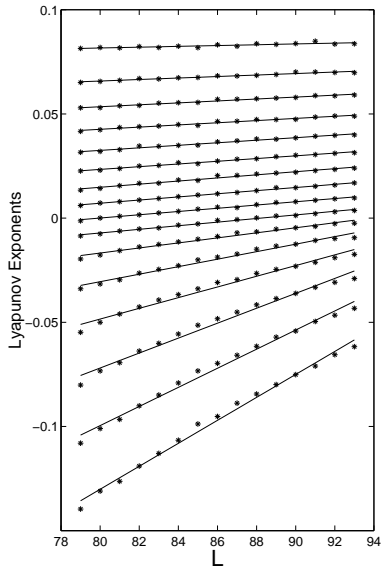}}
  \caption{
    The sixteen largest Lyapunov exponents~$\lambda_1 >
    \cdots > \lambda_{16}$ versus system length~$L$ for
    the same window of chaotic solutions discussed in
    Fig.~\ref{fig:Lyapunov_dimension}. For each value
    of~$L$, the $\lambda_i$ were estimated by
    estimating the asymptotic slope of the
    $\lambda_i(t)$ versus~$t$ curve for~$20,000 < t <
    200,000$. All Lyapunov exponents increase with~$L$
    in this regime except $\lambda_1$, which is
    approximately independent of~$L$ and so is an
    intensive quantity.  The lines through the points
    for a given~$\lambda_i$ are a guide to the eye and
    were determined by a least-squares fit. The
    exponents~$\lambda_i$ increases approximately, but
    not exactly, as linear functions of~$L$.}
  \label{fig:Lyapunov_exponents}
\end{figure}

Fig.~\ref{fig:Lyapunov_exponents} shows how the first
sixteen exponents~$\lambda_i$ vary with system size~$L$
over the same range of system sizes as
Fig.~\ref{fig:Lyapunov_dimension}. The largest
exponent~$\lambda_1$ is approximately independent
of~$L$ and so is an {\em intensive} quantity; an
increase of system size therefore does not change the
forecasting time~$1/\lambda_1$ over which the future
behavior of the field~$u$ becomes unpredictable,
despite the fact that the fractal dimension is becoming
correspondingly larger. The other exponents increase
roughly linearly with increasing~$L$, with the negative
exponents having the largest slope and showing
substantial deviations from a strictly linear
dependence on~$L$. Comparing
Fig.~\ref{fig:Lyapunov_exponents} with the Kaplan-Yorke
formula for the dimension
\begin{equation}
  \label{Eq-kaplan-yorke}
  D = K + {1 \over  |\lambda_{K+1}| } 
  \sum_{i=1}^K \lambda_i ,
\end{equation}
where~$K$ is the largest integer such that
$\sum_{i=1}^K\lambda_i \ge 0$, we see that the linear
behavior of~$D$ with~$L$ in
Fig.~\ref{fig:Lyapunov_dimension} is a subtle
consequence of how the~$\lambda_i$ vary with~$L$.

In conclusion, we have demonstrated the occurrence of
microextensive scaling of the Lyapunov fractal
dimension with system size for the one-dimensional
Kuramoto-Sivashinsky equation with rigid boundary
conditions. This suggests that a spatially extended
nonequilibrium dynamical system does not have to
increase its volume by some minimal amount for the
fractal dimension to increase.  Correspondingly, the
dimension length~$\xi_\delta$ does not have some direct
physical meaning as the characteristic size of a
dynamical subsystem, it is simply determined by the
linear growth of~$D$ with~$L$.  Our calculations
suggest two questions for further exploration. One is
whether there is a cutoff system size~$L_c$ above which
only chaotic solutions are found for the KS~eq. Second
is to understand mathematically how the geometry of a
strange attractor changes with system size~$L$ such
that that~$D$ varies exactly linearly with~$L$.

We thank Scott Zoldi for useful discussions and James
Gunton for informing us of his related calculations on
the Nicolaevski model.  This work was supported by NSF
grant DMS-9722814 and DOE grant DE-FG02-98ER14892.


\begin{references}
\bibitem{other-growth-rates} It is interesting to note
  that the number of positive Lyapunov exponents
  increases more slowly than the dimension~$D$ as $N_+
  = 0.14L - 1.5$, while the number of unstable modes
  increases more rapidly as~$N_u \approx
  0.32L$~\cite{Pomeau84}.
\end{references}

\end{document}